\documentstyle [12pt,epsfig]{article}
\begin{document}
\setlength{\baselineskip}{14pt}
\parindent=40 mm

\title{\Large\bf Chiral Lagrangians \\ and quark condensate in nuclei}
\author{ J. Delorme, G. Chanfray and M. Ericson\thanks{ Also at
 Theory Division, CERN,
CH-1211 Geneva
 23,
Switzerland} \\
{\small Institut de Physique Nucl\'{e}aire et IN2P3, CNRS,
Universit\'{e} Claude Bernard Lyon I,}
{\protect \vspace{-2mm}}
\and {\small 43 Bd. du 11 Novembre,
F-69622 Villeurbanne Cedex, France} \\ }

\date{ }

\maketitle

\begin{abstract}
{\normalsize
We study the evolution with density of
 the quark condensate in the nuclear medium
with interacting nucleons and including the short range correlations. We work
with two chiral
models, the linear sigma model and the non-linear one. For the last one we
 use two versions, one which does not satisfy PCAC, and another one which
does. We show that the quark condensate, as other observables,
 is independent on the variant selected. The application to 
physical pions excludes the linear sigma model as a
credible one. In the non-linear models restricted to pure s-wave pion-nucleon
scattering, our conclusions are the following: 

i) chiral symmetry does not imply a systematic reaction
 against symmetry restoration. In representations where PCAC holds, 
this absence results from a subtle cancellation.

ii) in the density evolution of the quark condensate, the two-pion
exchange potential, restricted to s-wave $\pi$N interaction, 
has a negligible effect, much smaller than that of the one-pion 
exchange. At variance with the linear sigma model, two-body 
contact terms simulating sigma meson exchange are not constrained by chiral
symmetry.
The same conclusion applies in the scattering of physical pions on nuclei, {\it
i.e.} to the s-wave optical potential.

iii)   On the other hand for
the effects of correlations calculable from chiral symmetry, those linked to 
the isospin symmetric amplitude are negligible.}
\end{abstract}
 
PACS numbers: 12.39.Fe 24.85.+p 25.80.Dj  

\begin{flushright}
LYCEN 9604 \\
CERN-TH/96-60 \\
nucl-th/9603005
\end{flushright}

\vspace{4mm}
\parindent=30pt
\section {\bf Introduction }

Nuclear matter produces a restoration of chiral symmetry which is
remarkably large. The quark condensate, which is
 the order parameter of the spontaneous symmetry breaking,
 decreases in magnitude by about 1/3 at the normal density
 $\rho_0$\cite{DL,CFG,LKW}. The bulk of this restoration
 comes from the additive
effect of independent nucleons. An effort has been made to study
 the restoration effect due to
the interaction between the nucleons \cite{CFG,CE,LK,BW}. In particular,
 Chanfray and Ericson \cite{CE} made the observation
that the quark condensate in a dense medium (of density $\rho$) is governed
 by the nuclear sigma commutator with the exact relation:
 \begin{equation}
\frac{< \overline {q}q (\rho) > } 
 {< \overline {q}q (0) > } =
1 - \frac{\rho\Sigma_A }{ f^2_\pi m^2_\pi}
\label{lin}  
\end{equation}
where $< \overline {q}q(\rho ) >$ is the quark condensate density in the
 medium and $< \overline {q}q(0) >$ its vacuum value. The pion mass and
decay constant are $m_\pi$ and $f_\pi$ ( = 94 MeV). We have denoted by
$\rho\Sigma_A $ the nuclear sigma commutator
 per unit volume with the definition for a piece of nuclear matter of
density $\rho$:
\begin{equation}
A\Sigma_A(\rho) = \langle \Psi_A\vert \big[Q_A,\big[Q_A,H\big]\big]
\vert \Psi_A \rangle
 \label{dcom}
\end{equation}  
where $\Psi_A$ is the nuclear ground state with total nucleon number $A$,
 $H$ the Hamiltonian and $Q_A$  the axial charge. In this way
the quantity $\Sigma_A(\rho)\>$, which is a function of the density,
is the effective sigma commutator per nucleon in
the nuclear medium. Thus the study of the in-medium
 quark condensate amounts to that of the sigma commutator. In the low
density limit, $\Sigma_A$ reduces to $\Sigma_N$  the nucleon sigma commutator
and the relation~(\ref{lin}) leads to the linear density dependence
 given in refs.~\cite{DL,CFG,LKW}. Notice that since $\Sigma_A$ is of order
$m_\pi^2$ the relative modification of the condensate survives in the chiral
limit $m_\pi \to 0\;$. 
 
Using the partially conserved axial current hypothesis (PCAC),
Chanfray and Ericson have related $\Sigma_A$ to the
soft pion amplitude on the nucleus. They have thus shown that a
correction to the additive assumption, {\it i.e.}
 $A\Sigma_A(\rho) = A\Sigma_N\,$, arises from
 the contribution of the pions exchanged between the nucleons which also
 participate in the restoration of chiral symmetry. This simple result does
 not depend on the original assumption that PCAC holds \cite{CE}.
 
Pushing the consequences of PCAC, M. Ericson has suggested the
existence of a reaction of
the nuclear medium against the restoration of chiral symmetry,
arising from the distortion
of the soft pion wave \cite{ME}.  This possibility was examined by Birse
and Mc Govern~\cite{BMG} in the linear
sigma model where PCAC holds. The distortion effect is indeed
 present but these authors have shown that it is overcompensated by exchange
 effects, linked to the $\sigma$
meson, such that there is an overall acceleration of the restoration instead
 of a hindrance. On the other hand T. Ericson has shown that the soft
 pion amplitude can be
appreciably affected by the correlations \cite{TE}, eventually cancelling,
as expected, the distortion effect in the limit of nucleons totally confined
inside the correlation hole.

The aim of the present work is twofold. We first want to complete
 the studies of refs.~\cite{ME,BMG,TE}
 by introducing the correlations in the linear sigma model
 in a consistent way,
which was not done previously. We will show that the effect of
the correlations should not
be discussed alone, but in connection with those of the $\sigma$ meson
 exchange introduced in
ref.~\cite{BMG}. Using the information provided by the scattering of
physical pions on nuclei we
will discuss the limitations of the linear sigma model and show
 that it cannot be trusted
for a quantitative discussion. The second and main aim of this
work is then to perform an
exploration of the nuclear sigma commutator in the non-linear sigma model,
 which does not assume the existence of a scalar meson. In the non-linear
 version the PCAC relation does
not automatically hold. This hypothesis is abandoned as a systematic assumption
and only the chiral properties of the
interaction are emphasized. However for certain representations it
 is possible to make a transformation on the field in such a way that PCAC is
 valid. In this case the nuclear
sigma commutator coincides with the soft pion amplitude and it
is possible to elucidate the interplay between the distortion and the exchange
 terms. We will show that physical
quantities, such as the condensate and the in-medium pion mass~\footnote{
An earlier comparison following different methods has been made
 between two non-linear Lagrangians and has reached the same conclusions about
 the effective pion mass~\cite{TW}.}
are independent of the
    representation, as expected from a general theorem \cite{CWZ}. We will
 also introduce the correlations in a
 consistent way in this model. We limit our discussion to the s-wave isospin
symmetric pion-nucleon amplitude.
\bigskip

\section{\bf Linear sigma model}
\bigskip
 
\begin{figure}
\begin{center}
\leavevmode
\epsfysize=10. cm
\epsfig{file=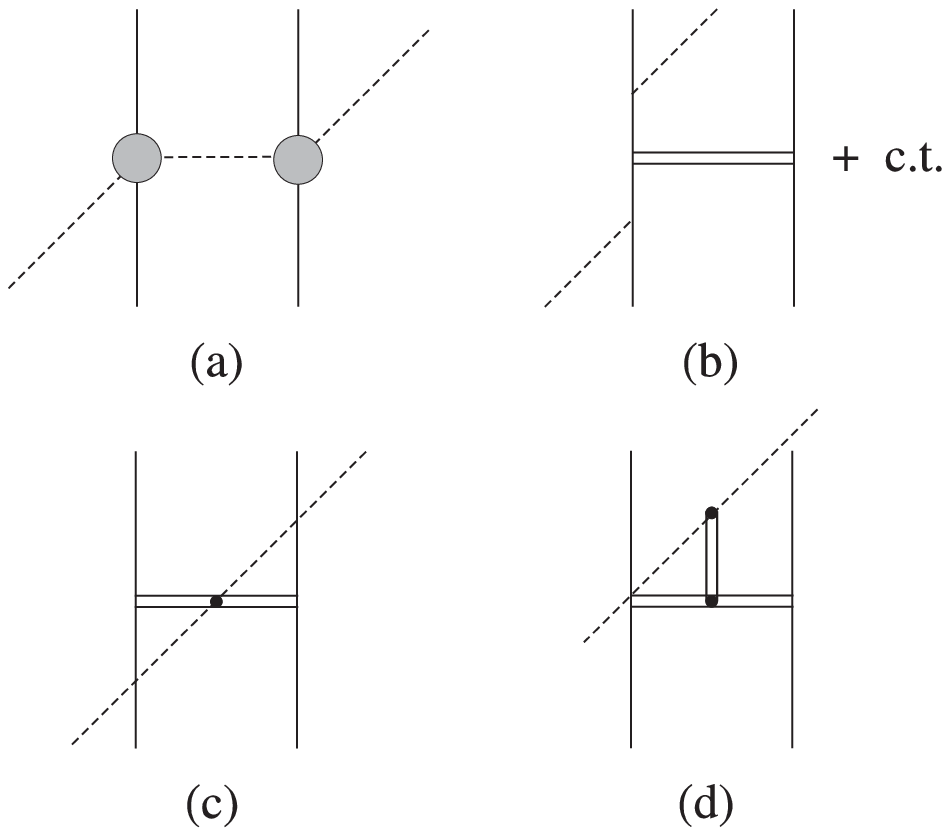}
\caption{ Two-body contributions to soft pion scattering in the linear sigma
model. Graph (a) is the coherent rescattering term where the round boxes
 represent the pion-nucleon scattering amplitude. Exchange of the $\sigma$
meson (double lines) gives rise to graphs (b) to (d).}
\label{figa}
\end{center}
\end{figure}

\begin{figure}
\begin{center}
\leavevmode
\epsfysize=10. cm
\epsfig{file=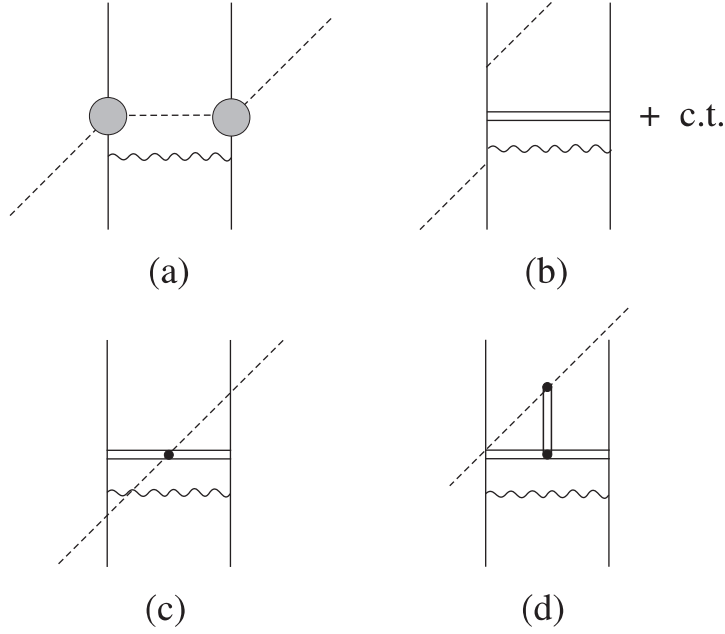}
\caption{ Same as for fig.~\protect\ref{figa} with wiggled lines standing for
 short-range correlations. Here graph (a) represents incoherent rescattering.}
\label{figb}
\end{center}
\end{figure}

We work in the linear sigma model, where the sigma commutator is
proportional to the soft pion amplitude. 
 We want to evaluate the scattering amplitude for soft pion on a nucleus,
 taken for simplicity as a piece of nuclear matter of density $\rho$. 
We denote $T$ the amplitude per unit volume so that
 $T = \Sigma_A \rho/f_\pi^2\,$.
 We work up to second order in the
 density.  For simplicity we limit our calculation to the
 tree level. The graphs contributing to second order in
 the density to the soft pion scattering amplitude are represented in 
fig.~\ref{figa}a-d.
 The coherent rescattering term is represented in fig.~\ref{figa}a.
 The graphs linked
to $\sigma$ exchange introduced in ref.~\cite{BMG} are
  shown in the graphs of fig.~\ref{figa}b-d.  The
short range correlations add the contributions of graphs~\ref{figb}a-d,
 where the
 wiggled line may be interpreted as the iterated exchange of a heavy meson,
 such as the $\omega$ meson, responsible for
 the short range repulsion. We take for the correlation function the
convenient schematic form 
$\,G({\hbox{\boldmath $r$ \unboldmath}})
 = - \, j_0 (q_cr)$ (with $q_c = m_\omega$) 
introduced by Weise~\cite{WSE},
which has the simple Fourier transform $G({\hbox{\boldmath $q$ \unboldmath}}) =
- 2\pi^2  \, \delta (q - q_c)/q_c^2 \>$ .
\bigskip

Denoting $T_\sigma$ the sum of the sigma exchange contributions 
in the correlated medium we obtain in the Hartree approximation which we
 follow throughout this paper for the clarity of the argument:
\begin{equation}
T_\sigma  = T_b + T_c + T_d =
\frac{g^2 \rho^2}{ f_\pi^2} \; \left[ 
\frac{5}{ 2} \; \frac{m^2_\pi }{ m^4_\sigma} - \frac{5}{ 2} \;
    \frac{m^2_\pi }{ (m^2_\sigma + q^2_c)^2} \;
- \frac{q^2_c }{ (m^2_\sigma + q^2_c)^2}\right] + O(m_\pi^4)\; , 
\label{sigex}
\end{equation}
where $g$ is the meson-nucleon coupling constant.
On the other hand, the sum of the coherent and incoherent rescattering is
(fig.~\ref{figa}a and~\ref{figb}a):
\begin{equation}
T_{resc.} =  
       \frac{g^2\rho^2  }{ f^2_\pi}\; \left[
      -  \frac{m^2_\pi}{ m_\sigma^4}\;
 +  \frac{m^2_\pi }{ (m^2_\sigma + q^2_c)^2} \;
    + \frac{q^2_c }{ (m^2_\sigma + q^2_c)^2}\right] + O(m_\pi^4)\; . 
\label{sigres}
\end{equation}
Both terms $T_\sigma$ and $T_{resc.}$
 are strongly affected by the correlations,
as apparent from the presence of a term proportional to $q_c^2$ which does
not vanish in the chiral limit $m_\pi \to 0\,$. However, when the sum of the
second order contributions is performed, the two terms
 in $q_c^2$ exactly cancel each
other. What this cancellation reveals is that the
 role of the incoherent rescattering (graph~\ref{figb}a) 
should not be discussed alone, otherwise chiral symmetry would be violated
 since the amplitude $T$ would not vanish in the chiral limit.
 Instead incoherent rescattering should be associated with the
contribution of the correlations in the $\sigma$ meson exchange
 graphs~\ref{figb}b-d so as to make the
soft amplitude go as $m_\pi^2\,$. This cancellation has not been discussed
 in the previous works of refs.~\cite{BMG,TE} but it is necessary
 to reach a consistent description. 

Adding the contribution of first order in the density to
 obtain the total amplitude we arrive at the following expression for the
density evolution of the quark condensate : 
\begin{equation}
  \frac{< \overline {q}q (\rho) > }{ 
 < \overline {q}q (0) > } -1 =  - \frac{T}{ m_\pi^2} 
= - \frac{\rho\Sigma_N  }{ f^2_\pi m_\pi^2} - \frac{3}{ 2} \;
    \frac{\rho^2\Sigma^2_N }{  f^4_\pi m^4_\pi} \> 
    \left[ 1 - \frac{1}{ \left ( 1 +   \frac
    {\displaystyle q^2_c }{  \displaystyle m^2_\sigma} \right)^2 } 
    \right]
    \left ( 1 - \frac{m^2_\pi }{ m^2_\sigma }\right)\> .
\label{consig}
\end{equation}
 
\noindent
Here we have introduced the sigma commutator of the
 nucleon, $\Sigma_N\>$ which in the linear sigma model writes: 
 $\Sigma_N = f_\pi g \> m^2_\pi / m^2_\sigma\>$ .
 Notice that $\Sigma_A$ could also be
 obtained directly as the expectation value of the symmetry breaking piece of
 the Lagrangian, i.e. of $f_\pi m^2_\pi \sigma \>$,
  in a correlated medium.
 The toy model of Ericson~ \cite{TE} corresponds to the limit of a large
 correlation hole compared to
 the extension of the condensate. In the linear sigma model this corresponds
to the condition $q_c \ll m_\sigma\,$. One sees on eq.~(\ref{consig}) that
 the term in $\rho^2$ 
then goes to zero and $\Sigma_A = \Sigma_N\>$ . The meson exchange terms
 disappear as expected, but also the
incoherent rescattering cancels the distortion, as found in ref.~\cite{TE}.
 In this case the
 quark condensate density in the nuclear medium evolves linearly with
 density, as given by the eq.~(\ref{lin}) if we replace
 $\Sigma_A$ by $\Sigma_N\>$.

This is an extreme situation and the question
 is how close it is to a realistic
one. In fact the size of the condensate, or equivalently the scalar radius of
the nucleon, has been extracted from experimental data~\cite{GLS1}.
 Its value, $r_s \approx 1.26\>$ fm is
 sensibly larger than the typical correlation
radius $r_c \approx 0.6\>$ fm. There is indeed a large overlap of the nucleon 
condensates which is neither reproduced in the linear sigma model nor in
 the toy model of ref.~\cite{TE}.

  We will discuss below in more detail the
limitations of the linear sigma model
 for a proper description of the problems of the quark condensate and  physical
 pion scattering. We first give in table~\ref{tab1} the value of the quark
condensate in this model at nuclear matter density $\rho_0 = 0.17$ fm$^{-3}$.
For the numerical evaluation we 
 take the parameters of the linear sigma model in such a way as to reproduce
 the experimental value of the sigma commutator: $\Sigma_N =$ 45
MeV~\cite{GLS2}. This
 fixes the $\sigma$ mass at $m_\sigma$ = 740~MeV. 
On the other hand we take for the cut-off parameter $q_c = m_\omega$ = 782~MeV. 
The acceleration of chiral symmetry restoration by $\sigma$
exchange terms is appreciable, 0.54 instead of 0.65 in the linear
 approximation eq.~(\ref{lin}).  
 
\begin{figure}
\begin{center}
\leavevmode
\epsfysize=12. cm
\epsfig{file=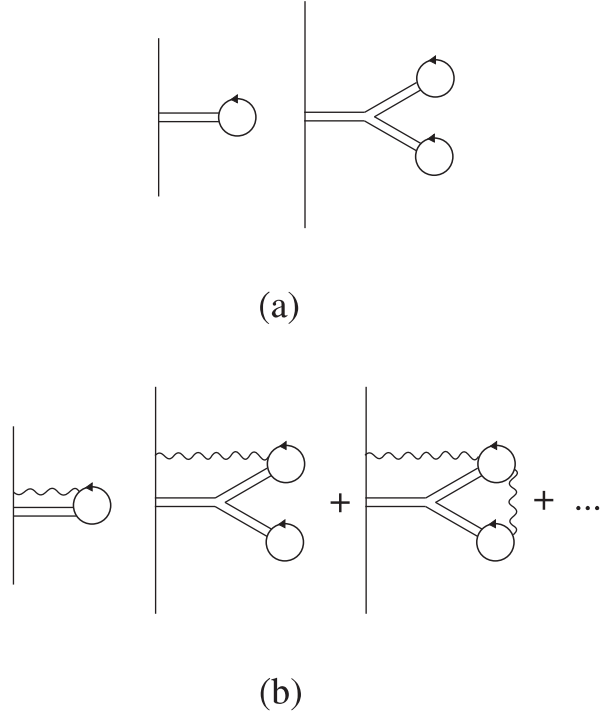}
\caption{ Contributions to the effective nucleon mass up to second order in
density in the linear sigma model without (a) and with (b) correlations.
At the order $\rho^2$ there are three graphs with one correlation function,
three with two and one with three (not shown).} 
\label{figc}
\end{center}
\end{figure}

 Since the linear sigma model is a consistent one which fulfils
 the requirements of chiral symmetry we find of interest to give also in 
table~\ref{tab1} the
 effective nucleon mass $M^{\ast}$ as evaluated from the graphs of
 fig.~\ref{figc}. Contrary to a common belief it does not follow exacly
 the same evolution as the condensate even to
first order in the density. Correlations do affect the first order in the mass
but not in the condensate.    
\bigskip

  The linear sigma model has serious shortcomings when it comes to a
quantitative description of  physical quantities. We mention some of the
problems.

\noindent
 {\it i)} The physics we are discussing is intimately
 linked to the behaviour of the isospin 
symmetric $\pi$N s-wave amplitude. The sigma model
 grossly fails to reproduce the corresponding scattering length. The
measured value is slightly repulsive
 $a^+ = - 0.010 \; m_\pi^{-1}\;$\cite{EW} 
arising essentially from the 
 pseudovector (PV) nucleon pole terms. The model gives instead a much
 larger repulsion, $a^+ = - 0.059 \; m^{-1}_\pi\;$.
  
\noindent
{\it ii)} in this model, the range of the condensate is determined by
 the $\sigma$ mass, which the value of the sigma commutator fixes
 at 740~MeV. The r.m.s. radius of
the condensate follows, $r_s = \sqrt{6}/m_\sigma \approx 0.65 \>$ fm.
 As already mentioned the value deduced from experiments is
much larger  $r_s = 1.26 \>$ fm.  
With a small extension of the condensate, one is
 closer to the limit where the condensate remains inside the correlation hole.
 We have seen that in this limit the incoherent rescattering cancels the
 coherent one, which is not the case in a realistic situation. The linear
 sigma model thus overestimates the incoherent rescattering effect, and more
 generally overstresses the role of the correlations. 

\noindent
{\it iii)} the last point concerns the $\sigma$ meson exchange effects. 
Is their magnitude realistic~? 
We can confront them with the experimental value of
the s-wave optical potential. Since the $\sigma$ exchange graphs of 
fig.~\ref{figa}c-d are sea-gull terms, they are
 independent of the energy, thus keeping the same value
for physical pions. This implies the existence
of a many-body correction due to $\sigma$ exchange in the s-wave
 optical potential. The latter is commonly parametrized as~\cite{EE}:
\begin{equation}
2\> m_\pi V_{opt} = - \left ( 1 + \frac{m_\pi}{ M} \right )
4\pi (b_0)_{\rm eff} \; \rho\; .
\label{vopt}
\end{equation}
 Since the quantity $(b_0)_{\rm eff}$     
incorporates many-body effects of various sources, it is a
 function of the density. The value of
$(b_0)_{\rm eff}$  deduced from the $\pi$-mesic data is   
$(b_0)_{\rm eff} = -0.029 \; m^{-1}_\pi\>$~\cite{EW}.
 The typical density explored by these data is somewhat
 lower than the nuclear matter one~\cite{SM}, {\it i.e.}
 $\rho \approx 0.7 \;\rho_0\>$. At this
 density, the linear sigma model predicts an exchange contribution 
(eq.~(\ref{sigex}) with the last term in $q^2_c$ removed) 
$ \Delta_{\sigma\, {\rm exch}} = - 0.023 \; m^{-1}_\pi\>$.
This would mean that most of the observed potential would have this
origin. Let us discuss
 this point in more detail.\par

In the low density limit the parameter $(b_0)_{\rm eff}$ should reduce to the
free nucleon value
$b_0 = - 0.010 \; m^{-1}_\pi\;$. 
 At finite density, there is one correction of well known origin. It is the
 rescattering with charge exchange, linked to the Pauli
 correlations~\cite{EE}. At $\rho = 0.7\rho_0$ 
 this adds $\Delta_1b = - 0.012\; m^{-1}_\pi\>$. 
We have also to consider the effect of correlations, $\Delta_2b$ linked 
to the isospin symmetric amplitude , i.e. the incoherent rescattering term
previously discussed and contained in eq.~(\ref{sigres}). 
This is not a sea-gull term. Therefore it depends on the energy, but
its dependence is mild and we neglect it. 
Removing the term in $q_c$ and the coherent rescattering in
eq.~(\ref{sigres}), we find:
\begin{equation}
\Delta_2b =
- \displaystyle{\frac{1}{ 4\pi (1 + \frac{m_\pi }{M})}} \;
 \frac{\rho\Sigma_N^2  }{ f^4_\pi m^2_\pi} \;
 \left ( 1 + 
\displaystyle{\frac{ q_c^2 }{  m^2_\sigma}}\right)^{-2} 
= - 0.003 \; m^{-1}_\pi\; ,
\label{d2b}
\end{equation} 
a small contribution.
Finally we have to consider also the exchange correction due to pion excess.
It represents the s-wave scattering on exchanged pions discussed by Chanfray
and Ericson in the case of soft pions~\cite{CE}.
 This is also a sea-gull term which
 keeps the same value for physical ones. With an excess number of pions per
 nucleon of 0.1 the corresponding
contribution is  $\Delta_3b = -0.004 \; m^{-1}_\pi\;$.\par
 
 Adding all the known contributions $b_0 + \Delta_1b + \Delta_2b + \Delta_3b$
we arrive already at  $(b_0)_{\rm eff} = - 0.029\; m^{-1}_\pi \> $, exactly 
 the experimental value.  Unless there is a large
attractive correction (but we do not see from where) there is no room for the
large $\sigma$ meson exchange correction implied by the linear 
sigma model, which is therefore unrealistic for $\pi$-nuclear scattering
problem as well.
  
Thus we can safely conclude that the meson exchange terms in the
scattering amplitude of soft, or physical pions, linked to the exchange of
 a scalar-isoscalar object, cannot be as large as given by the linear sigma
 model. This was our motivation to investigate the non-linear sigma model.
\bigskip
\noindent
\section{ \bf Non-linear sigma model}
\bigskip
Though non-linear sigma models have an already long story, their systematic 
application to nuclear forces and pion nuclear interactions has been
 recently popularized by Weinberg~\cite{WBGF}. We follow the guideline
 provided by an effective Lagrangian introduced in this spirit by Lynn
 in his study of nuclear matter~\cite{LYN}. 
We start from the linear sigma model with the Lagrangian density ${\cal L} =
{\cal L}_0 + {\cal L}_{CSB}$ where the chiral symmetric part
 writes in standard notations:
\begin{equation}
{\cal L}_0 = i\overline\psi\gamma^\mu\partial_\mu\psi
 - g\;\overline\psi(\sigma + i {\hbox{\boldmath $\tau . \pi$
\unboldmath}} \gamma_5)\psi
 + \frac{1}{ 2}( \partial_\mu\sigma \partial^\mu\sigma+ \partial_\mu
{\hbox{\boldmath $\pi$ \unboldmath}}.
\partial^\mu{\hbox{\boldmath$ \pi$\unboldmath}})
- \frac{\lambda}{ 4}(\sigma^2 +{\hbox {\boldmath $\pi$\unboldmath}}^2
 - v^2)^2\; , \label{lsig}
\end{equation}
where the limit $\lambda \to \infty$ is to be taken to get the non-linear
version. In order to recover the non-linear model of Lynn, we modify
the chiral symmetry breaking
term by adding a term bilinear in the nucleon field with the same
 chiral transformation
property as the usual term linear in the sigma field:  
 \begin{equation}
{\cal L}^{\sigma}_{CSB} = f_\pi m_\pi^2\sigma - \frac{\Sigma_N}{ f_\pi^2}
\sigma\;  \overline\psi(\sigma + i{\hbox {\boldmath $\tau . \pi$\unboldmath}}
\gamma_5)\psi\; ,
\label{lsigSB}
\end{equation} 
  with, as before, $\Sigma_N$ = 45 MeV the nucleon sigma commutator.

 We use the
 standard procedure to go to the non-linear version, writing $\sigma = f_\pi\>
\cos( F)$ and ${\hbox{\boldmath $\pi$\unboldmath}} = f_\pi\>
\hat{\hbox{\boldmath $\phi$\unboldmath}}\, \sin (F)$
 where $F$ is an arbitrary polynomial in $x = \phi/f_\pi\>$,
 which contains only odd powers of $x$.
We also transform the fermion field according to $\psi = S^\dagger N$ where
 $S^\dagger = \sqrt{(\sigma - i {\hbox{\boldmath $\tau . \pi$\unboldmath}}
\gamma_5)/f_\pi\;}$.
 A convenient choice for the function $F$, introduced by Weinberg is 
 $\sin (F) = x/(1+x^2/4)\;$~\cite{WBG}.
 The  corresponding chiral symmetry breaking Lagrangian is:
 
 \begin{equation}
{\cal L}^W_{CSB} = - \frac{1}{ 2}\;m_\pi^2\;\frac{{\hbox{\boldmath $\phi$
\unboldmath}}^2}
{(1+{\hbox{\boldmath $\phi$\unboldmath}}^2/4f_\pi^2)}
\left( 1 - \frac{\Sigma_N \bar N N}{ f_\pi^2 m_\pi^2}\right)\; .
\label{lwSB}
\end{equation}

In this representation the divergence of the axial current
does not satisfy PCAC since one has:
 \begin{equation}
\partial_\mu {\hbox{\boldmath$A$\unboldmath}}^\mu = f_\pi\;m_\pi^2\;
\frac{{\hbox{\boldmath $\phi$\unboldmath}}}{(1+{\hbox{\boldmath
 $\phi$\unboldmath}}^2/4f_\pi^2)}
\left( 1 - \frac{\Sigma_N \bar N N}{ f_\pi^2 m_\pi^2}\right)\; .\label{PCAC}
\end{equation}

\begin{figure}
\begin{center}
\leavevmode
\epsfysize=4.5 cm
\epsfig{file=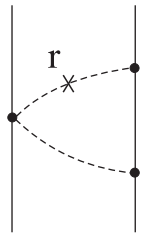}
\caption{ Interference between s- and p-wave pion nucleon couplings not
considered in the evaluation of the $\sigma$-term~ (\protect\ref{sigop}).
 The cross
marked with an $r$ denotes the place where the operator $\sigma_{op}$ acts.} 
\label{figd}
\end{center}
\end{figure}

\begin{figure}
\begin{center}
\leavevmode
\epsfysize=4.5 cm
\epsfig{file=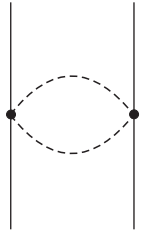}
\caption{  Two-pion exchange potential linked to the s-wave contact term 
 $V_{ct}^{2\pi}\,$(\protect\ref{v2pi}).}
\label{fige}
\end{center}
\end{figure}

The commutator of the axial charge operator with its time derivative, 
{\it i.e.} the $\sigma$-term which is a measure of explicit symmetry breaking
 reads: 
 \begin{eqnarray}
 \int d{\hbox{\boldmath $r$\unboldmath}}\;\sigma_{op}(r) & =
 & \big[Q_A,\big[Q_A,H\big]\big]
 \nonumber \\  
              & = &\frac{1}{ 2}\;m_\pi^2 f_\pi^2\; \int
 d{\hbox{\boldmath $r$\unboldmath}}\;
\left(\frac{{\hbox{\boldmath $\phi$\unboldmath}}^2(r)/f_\pi^2}
{(1+{\hbox{\boldmath $\phi$\unboldmath}}^2(r)/4f_\pi^2)}
      - 2\right)  \left(1 - \frac{\Sigma_N \bar N N(r)}
{ f_\pi^2 m_\pi^2}\right)\> .
\label{sigop}
\end{eqnarray} 
The variation of the condensate can be evaluated from the expectation value of
this operator, $\rho\Sigma_A(\rho)$, per unit volume of nuclear matter
(cf. eq.~\ref{lin}) divided by the squared pion mass. As already mentioned in
 the introduction, this does not mean that the medium dependence has to
 disappear in the chiral limit since $\Sigma_A/m_\pi^2$ is finite.
 The situation is very similar to that 
encountered in the Gell-Mann-Oakes-Renner relation
where the pion decay constant $f_\pi$ which is non zero in the symmetry limit
is related to the vacuum expectation value of the symmetry breaking
 Hamiltonian.   
To lowest order in the density the value of $\Sigma_A$ is that
of the free nucleon, $\Sigma_N\>$. We use perturbation theory to
evaluate the second order terms in the density. We retain only in
eq.~(\ref{sigop}) the
 lowest order terms in the pion field. The expectation
 value of the part $\frac{1}{ 2} m_\pi^2{\hbox{\boldmath $\phi$\unboldmath}}^2$
  of $\sigma_{op}$
 has been discussed by Chanfray and Ericson~\cite{CE}, restricting the
interaction hamiltonian to the pseudovector  $\pi$NN ( and also
   $\pi$N$\Delta$) coupling. They
showed that this contribution to the sigma commutator is
nothing else than the sigma
commutator of the pion excess. We extend here their calculation
 to incorporate the s-wave
contact term  $-\frac{1}{ 2}\;(\Sigma_N / f_\pi^2)\bar N N
 {\hbox{\boldmath $\phi$\unboldmath}}^2$
of the interaction Hamiltonian. There are interference terms between the
s-wave and p-wave operators, as represented in the graph of
fig.~\ref{figd}. We ignore them in
the present work. Our result for uniform nuclear matter, in the Hartree
approximation, is expressed as the following integral:
 
\begin{equation}
\Delta\Sigma_1^{2\pi} = -\frac{3}{ 2} \frac{\rho^2\Sigma_N^2 m_\pi^2}
{ f_\pi^4} \; \int \frac{(-i)d^4k}{ (2\pi)^4 (k^2-m_\pi^2)^3} 
          =  \frac{3}{ 2}\frac{\rho^2\Sigma_N^2}{ f_\pi^4} \;
 \frac{1}{ (4\pi)^2}\;
\int_{4m_\pi^2}^\infty\frac{m_\pi^2 dt}{ t^2(1-4m_\pi^2/t)^{-1/2}}\; .
\label{ds1}
\end{equation} 
 We have introduced the dispersive form in the right hand side of the
 second equality in order to
facilitate the comparison with the Feynman-Hellmann theorem. 
 The physical interpretation of this term is the same as the one
 given in ref.~\cite{CE}. It is the contribution of the excess pions linked
 to the s-wave contact term, as represented in
the graph~\ref{fige}.

Another contribution arises from the term $-\frac{1}{ 2}\>
(\Sigma_N/ f_\pi^2)\bar N N {\hbox{\boldmath $\phi$\unboldmath}}^2\>$
 of $\sigma_{op}\>$. It leads to a logarithmically divergent integral:
 
 \begin{equation}
\Delta\Sigma_2^{2\pi} = -\frac{3}{ 2} \frac{\rho^2\Sigma_N^2 }{ f_\pi^4} 
           \int \frac{(-i)d^4k}{ (2\pi)^4 (k^2-m_\pi^2)^2}\; 
       = - \frac{3}{ 2} \frac{\rho^2\Sigma_N^2}{ f_\pi^4} \;
 \frac{1}{ (4\pi)^2}\;
\int_{4m_\pi^2}^\infty \frac{(1-4m_\pi^2/t)^{1/2} dt }{ t}\; .
 \label{ds2}
\end{equation}
 The sum of these two contributions could also be obtained by
the Feynman-Hellmann theorem 
from the expectation value of the part of the two pion exchange
 potential linked to the s-wave contact term:
 
 \begin{equation}
V_{ct}^{2\pi}(r) = - \frac{3}{ 2}\frac{\Sigma_N^2}{ f_\pi^4(4\pi)^3}
 \int_{4m_\pi^2}^\infty (1-4m_\pi^2/t)^{1/2}
\frac{e^{-\sqrt t\, r}}{ r} dt\; .
\label{v2pi}
\end{equation}
 
We have to derive this expression with respect to the squared pion mass.
 The derivation of the integrand leads
 to the pion excess contribution discussed
 previously eq.~(\ref{ds1}). It adds a very small relative correction
 to $\Sigma_N\>$:
 $\Delta\Sigma_1^{2\pi}/\Sigma_N \rho = 3.6~10^{-3}\;$, which is not surprising
since the s-wave term~(\ref{v2pi}) is only a weak component of the two pion
exchange potential.  A second piece
arises from the derivation of the quantity $\Sigma_N^2\>$, $\Sigma_N$ being
proportional to $m_\pi^2\>$; it leads
to the previous diverging integral eq.~(\ref{ds2}). In order to evaluate
 it we introduce a cut-off mass  $\Lambda$  = 1 GeV.
 The corresponding value of $\Delta\Sigma_2^{2\pi}$ is again
small (a relative correction of $ - 1.4~10^{-2}$ to $\Sigma_N \rho \>$). 

\bigskip
 
Summarizing the results obtained with the Weinberg type Lagrangian,
 the evolution of the quark condensate with the density
 is mostly given, up to the normal
 density, by the linear term in the density:
\begin{equation}
\frac{< \overline {q}q (\rho) > }{ 
 < \overline {q}q (0) > } =
1 - \frac{\rho\Sigma_N }{ f^2_\pi m^2_\pi}\; .
\label{linN}
\end{equation}  
  The higher order corrections arise essentially from the modification
 of the pion number in the nucleus, which is expected to produce at
 normal density a mild  acceleration of the
restoration process (at density $\rho_0$, $\Sigma_A$ differs from $\Sigma_N$
  by $\approx$ 10\%).
 There is no reaction against the restoration, contrary to what can be
expected from PCAC.  
In the next paragraph we go to a representation
 which possesses PCAC, so as to elucidate the role of the distortion.

\bigskip
 The PCAC representation corresponds to $\sin (F) = x$, with a
subsequent transformation of the pion field:
$$
{\hbox{\boldmath $\varphi$\unboldmath}} = {\hbox{\boldmath $\phi$\unboldmath}}
\left( 1 - \frac{\Sigma_N \bar N N}{ f_\pi^2 m_\pi^2}\right)\> .$$ 
The chiral symmetry breaking lagrangian, expressed in the new
variable ${\hbox{\boldmath $\varphi$\unboldmath}}$ reads:
 
 \begin{eqnarray}
{\cal L}' _{CSB} & = & f_\pi^2m_\pi^2 \left( 1 - 
                           \frac{\Sigma_N \bar N N}{ f_\pi^2 m_\pi^2}\right)
\left[1 - \frac{{\hbox{\boldmath $\varphi$\unboldmath}}^2/f_\pi^2 }
{ (1 - \Sigma_N\bar N N/ f_\pi^2
m_\pi^2)^2}\right]^{1/2} \nonumber \\  
& \approx & f_\pi^2m_\pi^2 \left( 1 - 
       \frac  {\Sigma_N \bar N N}{ f_\pi^2 m_\pi^2}\right)
-\frac{1}{ 2}\;m_\pi^2\;\frac{{\hbox{\boldmath $\varphi$\unboldmath}}^2}
{ (1 - \Sigma_N\bar N N/ f_\pi^2 m_\pi^2)} + O(\varphi^4) \;   .  
\label{lPCAC}
\end{eqnarray}
 
Note that in this representation
the contact term $- \frac{1}{ 2}\;(\Sigma_N/ f_\pi^2)
 {\hbox{\boldmath $\varphi$\unboldmath}}^2\bar N N\>$,
 and hence also the soft pion amplitude on
the nucleon, has the opposite sign as compared to the previous Weinberg type 
representation. However the physical amplitude is the
same as will be shown below. The PCAC
relation which is now valid, allows the calculation of the nuclear sigma
 commutator as a
soft pion amplitude. If we ignore for the moment the one-pion and
 two-pion exchange contributions, the soft amplitude per unit volume is in the
Born approximation:
 \begin{equation}
T^{Born}_s = \frac{\rho\Sigma_N/f_\pi^2 }{ (1-\rho\Sigma_N/f_\pi^2m_\pi^2)}
\label{tbsoft}
\end{equation}
as follows readily from the interaction part of the Lagrangian
eq.~(\ref{lPCAC}). 
It displays a renormalization by higher order terms in the density,
 similar, but not equal, to the $\sigma$ meson exchange terms of Birse and Mc
Govern~\cite{BMG}.  Here they appear from contact terms in the Lagrangian
since the $\sigma$ meson mass is sent to infinity in the non-linear
version. The distortion
of the soft pion wave modifies the Born amplitude in such a way that
 the free nucleon amplitude is recovered:
 \begin{equation}
T_s = \frac{T^{Born}_s}{(1 + T^{Born}_s/m_\pi^2)} =
 \frac{\rho\Sigma_N}{ f_\pi^2}\; . 
\label{tsoft}
\end{equation}
 The soft pion amplitude per nucleon and hence 
 the sigma commutator $\Sigma_A$ is the
same as the free one. Here the distortion is exactly cancelled by
 the contact term, contrary
to what happens in the linear sigma model where it is overcompensated.
  The expression of the quark condensate density is the same as in the Weinberg
  representation, as it should~\cite{CWZ}. The introduction of pion
exchange effects does not alter this result at least to second order
 in the density. Indeed the one-pion contribution is
 obviously the same. As for the two-pion exchange eqs.~(\ref{ds1})
 and (\ref{ds2}), the change of sign of the contact term does not influence
the result as it appears quadratically in $V_{ct}^{2\pi}\>$.    
It is illustrative to check the independence on the Lagrangian also on the
pion effective mass in the medium, $m^{\ast}_\pi$, which obeys the equation
\begin{equation}
m^{\ast 2}_\pi = m_\pi^2 + \Pi(m^{\ast}_\pi,0) 
\label{mstar}
\end{equation} 
 where  $\Pi(m^{\ast}_\pi,0)$ is the pion self energy, taken at the energy
  $\omega = m_\pi^{\ast}$ and at zero three momentum. 
To find this last quantity we need the off-shell $\pi$N amplitude. For
simplicity, we neglect the small nucleon pole terms which would not change the
validity of the argument. In the representation of Weinberg
 the amplitude keeps the same value, $-\Sigma_N/f_\pi^2\>$,
 off- and on-shell. In
this case the effective mass has a very simple linear density
dependence:
 \begin{equation}
\frac{m^{\ast 2}_\pi }{ m^2_\pi} = 1 -
             \frac{\rho\Sigma_N }{ f_\pi^2 m_\pi^2}\; .\label
{mstlin}
\end{equation}
 
 In the PCAC representation instead, the $\pi$N amplitude has a complex off
 shell behaviour. Indeed it gets also a contribution from the pion kinetic
 energy term, $\frac{1}{ 2} D_\mu\vec\phi. D^\mu\vec\phi$ which
 expressed in the new field $\vec\varphi$ variable writes:

\begin{eqnarray}
\frac{1}{ 2}\; D_\mu{\hbox{\boldmath $\phi$\unboldmath}}\> 
. D^\mu{\hbox{\boldmath $\phi$\unboldmath}} & \approx &
\frac{1}{ 2}\; \frac{\partial_\mu{\hbox{\boldmath $\varphi$\unboldmath}}.
 \partial^\mu{\hbox{\boldmath $\varphi$\unboldmath}}}{ (1 -
 \Sigma_N\bar N N/ f_\pi^2m_\pi^2)^2} + \frac{\Sigma_N}{ f_\pi^2m_\pi^2}\; 
\frac{{\hbox{\boldmath $\varphi$\unboldmath}}.
 \partial_\mu{\hbox{\boldmath $\varphi$\unboldmath}} \>
 \partial^\mu(\bar N N)}{ (1 -
 \Sigma_N\bar N N/ f_\pi^2m_\pi^2)^3}  \nonumber   \\
& & + \frac{\Sigma_N^2}{ 2f_\pi^4m_\pi^4}\;\frac{
{\hbox{\boldmath $\varphi$\unboldmath}}^2 \partial_\mu
(\bar N N)\partial^\mu(\bar N N)}{ (1 - \Sigma_N\bar N N/ f_\pi^2m_\pi^2)^4}
+ O(\varphi^4)\; ,
\label{pikin}
\end{eqnarray}
where the operators $D_\mu{\hbox{\boldmath $\phi$\unboldmath}} =
 \partial_\mu {\hbox{\boldmath $\phi$\unboldmath}}/(1+ {\hbox
{\boldmath $\phi$\unboldmath}}^2/4f_\pi^2)$ are
 covariant derivatives of the pion field.
The presence of the nucleon fields in this expression leads to new
 contributions to the $\pi$N amplitude which takes the form:
\begin{equation}
\bar t_{\pi N} = \frac{\Sigma_N}{ f_\pi^2}\;\left(1 -
 \frac{q^2 + q'^2 }{ m_\pi^2}\right) 
\label{toff}
\end{equation}
where $q$ and $q'$ are the initial and final pion four momenta ($q^2 \equiv
\omega^2-{\hbox{\boldmath $q$\unboldmath}}^2\>$) and the bar over
 $t_{\pi N}$ indicates that
 the PV nucleon poles have not been considered. This expression
 satisfies Adler consistency conditions, together with the well known sign
 change between the soft and Cheng-Dashen points. Notice that the physical
amplitude is $\bar t_{\pi N} = - \Sigma_N/f_\pi^2$ as in the Weinberg
representation. 
 The nuclear Born amplitude $T^{Born}$, which is also the pion self-energy
 $\Pi$, writes:  
 \begin{equation}
T^{Born}(q^2) \equiv \Pi(\omega, {\hbox{\boldmath $q$\unboldmath}})
 = m_\pi^2\;\frac{x}{ (1 - x)} - q^2\;\frac{2x-x^2}{ (1-x)^2} 
\label{tborn}
\end{equation}
 where $x = \rho\Sigma_N/f_\pi^2m_\pi^2\,$ . In this expression, as in
eq.~(\ref{tbsoft}),
 the first piece in $m_\pi^2$ derives from
the interaction part in the pion mass term of
 eq.~(\ref{lPCAC}). The second piece in $q^2$ arises in the same way
 from the interaction
part of the kinetic energy term (eq.~(\ref{pikin}), first term).

The effective mass then obeys the equation:
\begin{equation}
 m^{\ast 2}_\pi\;\left(1 + \frac{2x-x^2}{ (1-x)^2}\right) =
 \frac{m_\pi^2 }{ (1-x)} 
\label{mstn}
\end{equation} 
which leads to the same result as the Weinberg
 Lagrangian eq.~(\ref{mstlin}), with the linear density dependence of
 $m_\pi^{\ast 2}\;$.
 The mass is indeed independent on the representation as already known from
previous work~\cite{TW}.   

Notice however that
 the simple linear drop of eq.~(\ref{mstlin}) which could lead to s-wave pion
 condensation~\cite{BKR} does not occur in reality, because the non-linear
 Lagrangian as introduced previously is incomplete. It fails to
 reproduce the (nearly) vanishing observed of the non-Born $\pi$N
amplitude at threshold (see ref.~\cite{DEE} for a general low energy amplitude
 with this property).
 It is then necessary to introduce extra terms,
 which are chirally invariant, so as to produce the threshold cancellation. 
Following refs.~\cite{GSS,BKM} as already done in ref.~\cite{TW}, we add to
 the previous Lagrangian a piece $\Delta{\cal L}$:
\begin{equation}
\Delta{\cal L} = \left(\frac{c_2}{ f_\pi^2}\;(v_\mu D^\mu
{\hbox{\boldmath $\phi$\unboldmath}})^2
+ \frac{c_3}{ f_\pi^2}\;D_\mu{\hbox{\boldmath $\phi$\unboldmath}}\> 
. D^\mu{\hbox{\boldmath $\phi$\unboldmath}}\right)\bar N N\; ,
\label{deltal}
\end{equation}
 where $v_\mu$ is the nucleon four-velocity which reduces to (1,0,0,0) in the
nucleon rest frame. This extra piece does not change the condensate but it
modifies the effective mass:
\begin{equation}
\frac{m_\pi^{\ast 2}}{ m_\pi^2} = \frac{1 -
\rho\Sigma_N/f_\pi^2m_\pi^2}
{1 + 2\rho(c_2 + c_3)/f_\pi^2}\; .
\label{mstnn}
 \end{equation}  
The threshold cancellation of the non-Born amplitude implies $2(c_2+c_3)m_\pi^2
\approx - \Sigma_N$ such that $m_\pi^{\ast 2}/ m_\pi^2  \approx 1$. This
 relation was first derived on a multiple scattering
basis~\cite{DEE} and then in a chiral Lagrangian approach~\cite{TW} (see also
~\cite{AB,BR}).

We summarize the discussion of this section in table~\ref{tab2} where we 
compare the results obtained with the Weinberg and PCAC
Lagrangians for the amplitudes and the observables. We also give there the
expression of the inverse pion propagator
 $D^{-1}_\pi(\omega, {\hbox{\boldmath $q$\unboldmath}} \>)$ 
in the two cases. They are identical up to a factor $(1-x)^2$, with in
particular the same dispersion law. We recover here the expression of
 the propagators already derived along somewhat different lines
 by Thorsson and Wirzba~\cite{TW} in their
comparative study between two Lagrangians of Weinberg and PCAC types.    
We could have discussed also the behaviour of the nucleon effective
mass $M^{\ast}$. Retaining only the s-wave pion-nucleon interaction we would
get to first order in the density: $M^{\ast}/M = 1 + 
\Delta \Sigma_2^{2\pi}/\rho M +$ terms in $c_2, c_3\>$. It is seen that this
 is a very small correction which bears no resemblance with the evolution of
 the condensate~(\ref{linN}). In fact, in the effective Lagrangian
approach~\cite{WBGF},\cite{LYN} the nucleon effective mass is mainly governed
 by a contact term $\frac{1}{2} C_S^2\> (\bar \psi \psi)(\bar \psi
\psi)\>$ which is a chiral invariant and thus does not contribute to the
$\sigma$ term~(\ref{sigop}). We get then: $M^{\ast}/M = 1 - \rho C_S^2/M \>$
 with no relation to the condensate, contrary to what occurred, at least in the
absence of correlations, in the linear sigma model (see table~\ref{tab1}).
 The latter situation is
however quite peculiar in that the nucleon mass is generated by spontaneous
symmetry breaking. It may happen of course that, at the deeper level of QCD,
the phenomenological coefficients $M, C_S^2,\, \Sigma_N \cdots$ of the
 non-linear Lagrangians are connected so that the evolution of the nucleon
 effective mass and that of the quark condensate become similar.     

We will now discuss the influence of the correlations on the condensate. In the
 representation of Weinberg it is clear that the quark condensate
 cannot be affected by the incoherent rescattering of the
soft pions discussed in ref.~\cite{TE} since the nuclear sigma commutator
 bears no relation to the soft pion
amplitude. On the other hand it is interesting to understand in the
 representation in which PCAC holds how the effect of
 incoherent rescattering can be cancelled in the soft pion amplitude
which is the sigma commutator. This investigation is also
 interesting in connection with the scattering of
physical pions. Indeed the isospin symmetric optical potential
 is strongly influenced by
the correlations, as the corresponding  $\pi$N scattering length is
 very small. The incoherent rescattering effect linked to the charge-exchange
 amplitude has been discussed a long time ago~\cite{EE}. The one linked to the
 isospin-symmetric amplitude is more
debatable as the off-shell behaviour of this amplitude  is model
 dependent.  In the PCAC
representation it is given by expression~ (\ref{toff}). We will first
 discuss the role of the correlations on the soft pion amplitude
 and then we extend the result to the physical threshold pions.

\begin{figure}
\begin{center}
\leavevmode
\epsfxsize=9. cm
\epsfig{file=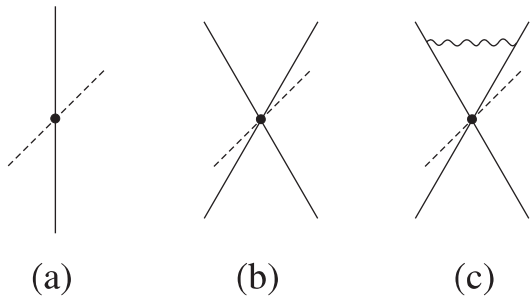}
\caption{ Soft pion scattering up to second order in density in a
 non-linear
 Lagrangian satisfying PCAC (\protect\ref{lsoft}). Graph (a) and (b) are one-
 and two-body contact terms with correlations absent in the latter.   
Correlations (wiggly line) are instead accounted for by graph (c).}
\label{figf}
\end{center}
\end{figure}

\begin{figure}
\begin{center}
\leavevmode
\epsfxsize=10. cm
\epsfig{file=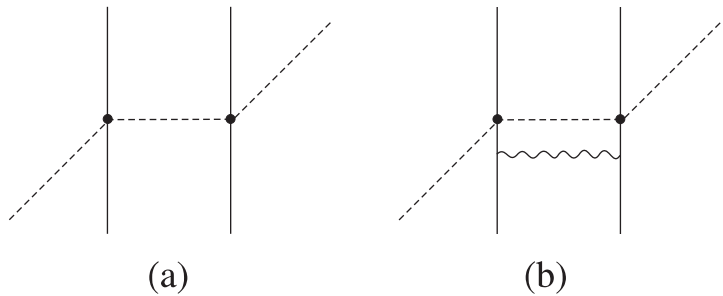}
\end{center}
\caption{ Same as fig.~\protect\ref{figf} but for the case of coherent (a)
 and incoherent (b) rescattering.}   
\label{figg}
\end{figure}

We limit our consideration to two-body correlations and hence we work to second
order in the density. To this order and in the soft limit the relevant pieces
 of the Lagrangian in the PCAC representation are:
\begin{eqnarray}
{\cal L}_{soft} & \approx & \frac{1}{ 2}\;\frac{\Sigma_N }{ f_\pi^2}\;
{\hbox{\boldmath $\varphi$\unboldmath}}^2\bar N N  \nonumber         \\
& &  +   \frac {1}{ 2}\;\frac{\Sigma_N^2 }{ f_\pi^4m_\pi^2}\;
{\hbox{\boldmath $\varphi$\unboldmath}}^2(\bar
N N)^2 - \frac{1}{ 2}\;\frac{\Sigma_N^2 }{ f_\pi^4m_\pi^4}\;{\hbox{\boldmath
$\varphi$\unboldmath}}^2 \partial_\mu(\bar
N N)\partial^\mu(\bar N N)\; .  
\label{lsoft}
\end{eqnarray}
In a correlated medium the amplitude can be evaluated from the
graphs of fig.~\ref{figf} and~\ref{figg}. Those of fig.~\ref{figf} are
 the contact terms while those of
fig.~\ref{figg} represent the coherent and incoherent rescattering. The latter
corresponds to the Lorentz-Lorenz effect well known in the p-wave case.
 In the absence of correlations and in the static approximation only the first
two pieces of the Lagrangian eq.~(\ref{lsoft}) contribute an amount
 (graphs~\ref{figf}a and~\ref{figf}b): 
\begin{equation}
T_1 = m_\pi^2 x(1+x)
\label{tct1}
\end{equation}
which is nothing else than the expansion of $x/(1-x)$ of
formula~(\ref{tbsoft}).
 With correlations the graph~\ref{figf}c gives:
\begin{equation}
T_2 = m_\pi^2 x^2\;\int \frac{d{\hbox{\boldmath $q$\unboldmath}}''}{ (2\pi)^3}
 (m_\pi^2+q\,''^2)G({\hbox{\boldmath $q$\unboldmath}}'')
\label{tctg}
\end{equation} 
where the two parts in the integrand arise from the second and third pieces of
the Lagrangian respectively. 
The rescattering terms ~\ref{figg}a and ~\ref{figg}b add:
\begin{equation}
T_3 = - m_\pi^2 x^2\;\left(1 +\int \frac{d{\hbox{\boldmath $q$\unboldmath}}''}
{ (2\pi)^3}  \frac {(m_\pi^2+q\,''^2)^2 }{
 (m_\pi^2+q\,''^2)}G({\hbox{\boldmath $q$\unboldmath}}'')\right)\> . 
 \label{tresc}
\end{equation}
Summing all the pieces the total soft pion amplitude  thus reduces to:
 \begin{equation}
T = T_1 + T_2 + T_3 = \rho\Sigma_N/f_\pi^2\; .
\label{ttot}
\end{equation}  
The overall effect of the correlations disappears due to cancellation between
the Lorentz-Lorenz term ({\it i.e.} the incoherent rescattering of
graph~\ref{figg}b)
and the correlation contribution in the contact terms. 
The soft amplitude and hence the sigma commutator are affected neither by
 distortion nor by correlations. Thus we recover the result of the Weinberg
type Lagrangian. The validity of our statement about correlations is however
limited to the simplest case where the chiral Lagrangians (e.g. our
 eqs.(\ref{lsig}-\ref{lwSB})) do not contain terms of second order or more
 in nucleon fields bilinears $\bar\psi \psi\>$. As we will be comment later on,
 correlations effects cannot be excuded on the sole basis of chiral symmetry
  for Lagrangians extended to higher order.
 
It is natural to extend our study to the case of threshold physical
pions, as we did in the linear sigma model. Our aim is first to understand if,
 in the non-linear model, chiral symmetry imposes the existence of
 two-body terms in the s-wave optical potential. This was the case in the linear
sigma model. The second point is to make a quantitative evaluation of the
influence of the correlations linked to the isospin symmetric $\pi$N amplitude.

Concerning the first point, we recall that in the PCAC representation and for
soft pions contact terms are needed to cancel the distortion. The question is
then: are these contact terms present in the s-wave optical potential as it is
the case in the linear sigma model~? In the Weinberg representation contact
terms are totally absent. Their cancellation is therefore plausible in the PCAC
representation. The problem is to understand the mechanism for cancellation.
For simplicity we ignore the extra terms in $c_2$ and $c_3$ which do not change
the essence of the discussion. 

In the PCAC representation the pion self energy has been given in
 eq.~(\ref{tborn}). It
displays a priori many-body terms. Threshold pions correspond to the situation
where $\omega = m_\pi, \; {\hbox{\boldmath $q$\unboldmath}} = 0$
 which are valid outside the
 nucleus. For the inside conditions we take the example of a spherical
 nucleus with a uniform density. In the interior of the nucleus the pion keeps
 its energy and acquires a momentum ${\hbox{\boldmath $q$\unboldmath}}_{eff}$
 solution of the dispersion equation:
\begin{equation} 
\left[ \omega^2 - {\hbox{\boldmath $q$\unboldmath}}_{eff}^2 - m_\pi^2 -
 \Pi(\omega, {\hbox{\boldmath $q$\unboldmath}}_{eff})
\right] _{\omega = m_\pi} = 0 \; , 
\label{disp}
\end{equation}
which gives ${\hbox{\boldmath $q$\unboldmath}}_{eff}^2 = m_\pi^2 x$ such that
 $q^2 = m_\pi^2(1-x)\;$. In the
usual expression of the potential $2m_\pi V_{opt} =
 - \left(1+\displaystyle{\frac{m_\pi}{ M}}\right) 4\pi b_0 \rho$ the
momentum dependence is not explicitly taken into account but it has to be
implicitly incorporated in the parameter $b_0$ as follows:
 \begin{equation}
-\left(1+\frac{m_\pi}{ M}\right) 4\pi b_0 \rho = \Pi(m_\pi, 
{\hbox{\boldmath $q$\unboldmath}}_{eff}) = m_\pi^2\left[\frac{x }{ (1-x)} -
(1-x)\frac{2x-x^2}{ (1-x)^2}\right] = - {\rho\Sigma_N }{ f_\pi^2}\; . 
\label{b0eff}
\end{equation}
This expression displays no two-body terms and is the same as for the Weinberg
Lagrangian. Thus in the PCAC representation of the non-linear sigma model the
momentum dependence of the self energy is such that many-body terms are absent
in the s-wave optical potential. In the linear model instead,
 the $\pi$N amplitude
 $\displaystyle{\bar t_{\pi N} = \frac{g}{ f_\pi}\;
\frac{m_\pi^2-t}{ m_\sigma^2-t}}$
  which also satisfies the Adler condition, depends only
 on the four momentum transfer variable $t$ so that in the forward direction
 there is no three momentum dependence. Hence the self energy is momentum
independent and there is no cancellation of the contact terms. In that
 respect the non-linear and linear sigma models are quite different. 

 We now turn to the influence of correlations on the s-wave pion-nucleus
 optical potential. For this purpose it is
simpler to use the Weinberg Lagrangian supplemented with the $\Delta{\cal L}$
 term introduced previously in eq.~(\ref{deltal}). Indeed in this case
 there are no two-body contact terms and correlations
 enter only as a Lorentz-Lorenz effect. For pions at rest with energy $\omega$
 the amplitude in the Born approximation ({\it i.e.} the pion self-energy)
takes the following form in a correlated medium:  
\begin{equation}
T^{Born} \equiv \Pi(\omega,0) = \rho\> t_{\pi N}(\omega,0)\;
      \left[ 1 - \rho\> t_{\pi N}(\omega,0) \;
 \int d{\hbox{\boldmath $q$\unboldmath}}''\frac{ G({\hbox{\boldmath $q$
\unboldmath}}'')}
{ (m_\pi^2 + q\,''^2)}\right] \> .  
\label{tphys}
\end{equation}
Here $t_{\pi N}(\omega,0)$ is the full off-shell $\pi$N amplitude including
    the nucleon poles in pseudovector coupling (first term on the right hand
side of the following equation):      
\begin{equation}
t_{\pi N}(\omega,0) = g^2\;\frac{\omega^2}{ 4M^3} -
\frac{\Sigma_N }{ f_\pi^2}\;\left(1+\frac{2(c_2+c_3)\omega^2}
{\Sigma_N}\right) \> .
 \label{ttos}
\end{equation}
 The s-wave pion optical potential is obtained from the self-energy
  eq.~(\ref{tphys}) taken at $\omega = m_\pi\,$:
 $2m_\pi V_{opt} = \Pi(m_\pi,0)\>$.
The pole terms can be ignored in the part with the correlation integral
 which gives a very small contribution. As for the remaining terms
the effect of the correlations is totally negligible due to the cancellation
which occurs at threshold  $\Sigma_N + 2(c_2+c_3)m_\pi^2 \approx 0\,$.
 The same conclusion would be obtained in a more complicated way with the PCAC
 representation. Thus we can conclude that only the incoherent rescattering
 with charge exchange, which has not been introduced in the present discussion,
  is relevant for the s-wave optical potential, at variance with the findings
 of Salcedo {\it et al.}~\cite{SHOS}. 

   One conclusion of our study
 with the non-linear sigma model is the absence of
two-body terms in the nuclear sigma commutator $\Sigma_A$ (apart from those
 due to pion exchange) and in the s-wave optical potential. The meaning of this
statement is that chiral symmetry does not impose them. They can nevertheless
be present. Indeed we have used the simplest forms of the chiral Lagrangians.
In the nuclear medium, at the same order in the chiral expansion one can add
 terms quadratic in the nucleon density or higher, either breaking chiral
 symmetry such as ${\hbox{\boldmath $\phi$\unboldmath}}^2
/(1+{\hbox{\boldmath $\phi$\unboldmath}}^2/4f_\pi^2)
(\bar \psi \psi)^2$ or conserving
 it as e.g. $D_\mu{\hbox{\boldmath $\phi$\unboldmath}}\,
. D^\mu{\hbox{\boldmath $\phi$\unboldmath}} (\bar \psi \psi)^2\>$. 
 In the spirit of chiral perturbation expansions all the
 coefficients are phenomenological and should be
determined by experiment though their order of magnitude can be guessed by
dimensional counting arguments~\cite{LYN,MG}. A priori they can be of any sign,
  accelerating or hindering chiral symmetry restoration. We are planning to
 explore which experiments can be informative on these terms,
 beyond the obvious one of pion-nucleus scattering.
In the same way nucleon-nucleon interaction terms built with squared bilinears 
$(\bar \psi\Gamma\psi)^2\;$ ($\Gamma \equiv \gamma$ matrices) should be
 present as introduced e.g. in refs~\cite{WBGF,LYN}
 to simulate heavy exchanges. With such Lagrangians, the effects of correlations
 can be buried in the coefficients of some contact terms or/and have to be
 explicitly computed as loop contributions. 

\section{\bf Conclusion}

To conclude the present study, we have investigated the evolution with
density of the quark condensate beyond the approximation of non interacting
nucleons. In the non-linear sigma model we have found that the calculable
correction to the linear dependence of the quark condensate arises
  from the one-pion and two-pion exchange potentials. The first
 one has been already discussed by Chanfray and Ericson~\cite{CE}.
 As for the latter we have evaluated it in the framework of the non
linear model and found it to be small. 
However we stress that this is only a part of the two-pion exchange potential,
the part linked to the contact s-wave $\pi$-nucleon amplitude.
The bulk of the potential arises from the p-wave coupling terms which is
outside the present approach. The evaluation of its role is in progress.
 Chiral symmetry does not impose other two-body
contributions contrary to what occurs in the linear sigma model. They can
nevertheless be present but have to be determined empirically.  
In particular there is no
systematic reaction against the restoration of chiral symmetry.      

Since the quark condensate is the order parameter of chiral symmetry, it is
intimately linked to the physics of the pion which is the Goldstone boson of
 this symmetry in the broken phase. It is thus natural to apply the concepts
developed for the condensate in the nuclear medium to the pion-nucleus
    interaction. The absence of two-body terms imposed by chiral symmetry
in the condensate reflects in their absence in the $\pi$-nucleus optical
potential. The possible ones have to be determined empirically. 
We have also found that correlations act only through the well known
incoherent rescattering with charge exchange. 

Thus the non-linear sigma model has been quite helpful to make clear several
 intriguing points. We have displayed on an example how the independence of the
condensate and other physical quantities on the particular form of the
Lagrangian is achieved. In the version where PCAC holds, this property occurs
through a subtle cancellation which eliminates both coherent (distortion) and
 incoherent rescattering (effect of correlations) in the soft pion amplitude.
We have made a quantitative estimate of the (small) contribution of the contact
two-pion exchange potential $V^{2\pi}_{ct}$ to the condensate, which can be
extended to the full two-pion exchange. Though PCAC may be misleading
in the sense that it introduces artificial effects eventually cancelling each
other, the derivation of the condensate properties from a soft pion amplitude
has the advantage of stimulating the application of the same methods to the
 domain of physical pions. Following this line we have been lead to question
the use of the linear sigma model for the estimate of heavy meson exchanges.   
Such cross-fertilization between physics related to QCD and pion physics
will likely bring new results in both fields.

\vspace{4mm}
\section{\bf Acknowledgement}

\vspace{4mm}
We are grateful to Drs. T. Ericson, M. Lutz and M. Rho
 for stimulating discussions.

\newpage 

\newpage
 
\begin{table}
\begin{center}
\caption{ Evolution of the quark condensate and the effective nucleon mass
to second order in the nuclear density in the linear sigma model. The third
term in the formula for $M^{\ast}/M$ is obtained after approximated treatment
of the graphs with 2 and 3 correlation functions. The numbers in third column
are given for density $\rho = \rho_0$.}
\label{tab1}
\medskip
\begin{tabular}
{|c||c|c|}  \hline
 &  &    \\
$\begin{array}{c} < \overline {q}q (\rho) > / < \overline {q}q (0) > \\
{\rm (condensate)}\end{array}$
&$1 - \displaystyle{\frac{\rho \Sigma_N }{ f^2_\pi m^2_\pi }} 
  - \displaystyle{\frac{3}{ 2} \; 
  \frac{\rho^2 \Sigma^2_N }{ f^4_\pi m^4_\pi}} 
  \left ( 
  1 -  \displaystyle{\frac{ 1}{
      \left( 1 +  q^2_c/ m_\sigma^2 \right)^2}} \right ) $ &\quad 0.54\quad \\
 &  & \\
  \hline 
 &  &  \\  
 & \hspace{- 2. cm} $1 - \displaystyle{\frac{\rho \Sigma_N }
                                                 { f^2_\pi m^2_\pi}} 
    \left ( 1 - \displaystyle{\frac{1}{ (1 + q^2_c/m^2_\sigma)}}\right )$
 & \\
$M^{\ast}/M$ &  & \quad 0.71 \quad \\ 
& \hfill
   $ - \displaystyle{\frac{3}{ 2} \frac{\rho^2 \Sigma^2_N }{ f^4_\pi m^4_\pi}}
    \left( 1 - 
     \displaystyle{\frac {1}{ ( 1 + q^2_c/ m^2_\sigma )^2 }} \right )$ & \\
& &  \\ \hline
\end{tabular}
\end{center}
\end{table}

\begin{table}
\begin{center}
\caption{Comparison of the results obtained with two non-linear Lagrangians,
 ${\cal L}^W$ which does not satisfies PCAC and ${\cal L}'$ which does.
The successive lines give the pion-nucleon amplitude, soft pion-nucleus
amplitude, pion self-energy (or Born pion-nucleus amplitude), inverse pion
propagator, squared pion effective mass and density
 evolution of the quark condensate. For simplicity the nucleon PV pole terms
are not included in the calculation.  We use $x =
\rho\Sigma_N/f_\pi^2 m_\pi^2$ as in the text and
 we define $x_{2,3} = \rho\, c_{2,3}/f_\pi^2\;$. }      
\label{tab2}
\medskip
\begin{tabular}
 {|c||c|c|}  \hline
& & \\
 & ${\cal L}^W$ &$ {\cal L}' $   \\ \hline \hline 
& & \\
$ \bar t_{\pi N}$ &$\displaystyle{-\frac{\Sigma_N}{ f_\pi^2} }$ &
 \hspace{-2. cm}$\displaystyle{\frac{\Sigma_N}{ f_\pi^2}\big(1 -
 \frac{q^2 + q'^2 }{ m_\pi^2}\big)} $ \\  
& & 
\hfill$
-\displaystyle{\frac{2\left(c_2\nu^2 + c_3 2M\nu_B\right)}{f_\pi^2}}$
 \\ 
 &  & \\
  \hline 
 &  &  \\  
$\displaystyle{ \frac{T_s}{m_\pi^2}}$ &$ - x/(1-x)$ 
& $\displaystyle{\frac{x/(1-x)}{1 + x/(1-x)}}
 = x $ \\
 &  & \\
  \hline 
 &  &  \\  
$\displaystyle{ \frac{\Pi(\omega,{\hbox{\boldmath $q$\unboldmath}} \,)}
{m_\pi^2}}$ & $ - x -
2(x_2+x_3)\displaystyle{\frac{\omega^2}{m_\pi^2}}$
& 
 $\displaystyle{ \frac{x}{(1-x)} -  
\frac{2(x+x_2+x_3)-x^2}{ (1-x)^2}\frac{\omega^2}{m_\pi^2}}$ \\ 
    &\hfill $+ 2x_3\displaystyle{\frac{{{\hbox{\boldmath $q$\unboldmath}}}^2}
{m_\pi^2}}$ & 
 \hfill $\displaystyle{ +\frac{2(x+x_3)-x^2}{(1-x)^2}\frac{
{\hbox{\boldmath $q$\unboldmath}}^2}{m_\pi^2}}$ \\ 
& & \\
\hline 
 &  &  \\ 
$ D_\pi ^{-1} (\omega, {\hbox{\boldmath $q$\unboldmath}} \,) $ &
 $\big [ 1+2(x_2+x_3) \big]\omega^2$
& $ \bigg[\big [1+2(x_2+x_3)\big]\omega^2
 - (1+2x_3){\hbox{\boldmath $q$\unboldmath}}^2$ \\
 &\hfill $ -  (1+2x_3){\hbox{\boldmath $q$\unboldmath}}
^2 - (1-x)m_\pi^2$ &\hfill $  -
 (1-x)m_\pi^2 \bigg] /(1-x)^2 $ \\ 
& & \\
\hline 
 &  &  \\ 
$\displaystyle{\frac{ m_\pi^{\ast 2}}{m_\pi^2}}$ &
 $\displaystyle{\frac{1-x}{1+2(x_2+x_3)}}$ & 
$ \displaystyle{\frac{1-x}{1+2(x_2+x_3)}} $\\
& & \\
\hline 
 &  &  \\ 
$\displaystyle{\frac{< \overline {q}q (\rho) >} { < \overline {q}q (0) >}}$
 & $1-x$ & $1-x $ \\
& &  \\ \hline
\end{tabular}
\end{center}
\end{table}
\end{document}